Synthesis and Structural details of $BiOCu_{1-x}S$: Possible new entrant in series of exotic superconductors?

Anand Pal, H. Kishan and V.P.S. Awana*,

(*: awana@mail.nplindia.ernet.in: www.freewebs.com/vpsawana/)

National Physical Laboratory (CSIR), Dr. K.S. Krishnan Marg, New Delhi-110012, India

We report an easy route single step synthesis of BiOCuS with and without Cu deficiency. The title compound is synthesized via solid state reaction route by encapsulation in an evacuated ($10^{-3}$ Torr) quartz tube. Mixed components of the ingredients in stoichiometric ratio ($1/3Bi_2O_3$ +0.34Bi +$Cu_{1-x}$ + S) are pelletized, sealed in evacuated quartz tube and heat treated for 30 hours at 500 $^0$C. Finally the sample is allowed to cool down to room temperature. The resultant compound is black is color and could not hold in pellet form, but is powdered. X-ray diffraction Reitveld analysis is carried out on all three samples of series $BiOCu_{1-x}S$ with x = 0.0, 0.10 and 0.15. These samples crystallize in single phase with space group P4/nmm and with cell parameters as a = 3.868 A and c = 8.557 A for stoichiometric BiOCuS. The volume of the cell slightly increases with an increase in Cu deficiency. The co-ordinate positions are determined by fitting the observed XRD patterns of the studied samples.

*: Corresponding Author

e-mail: awana@mail.nplindia.ernet.in: Fax/Phone – 0091-11-45609210/9310

Home Page: www.freewebs.com/vpsawana/

Very recent invention of superconductivity in Fe based compounds REFeAsO (RE=rare earth) [1] and $FeSe_{1-x}Te_x$ [2] has opened new channels and challenges to the solid state physics community. Worth mentioning is the fact that after the avalanche of high $T_c$ superconductivity in 1986 [3], although besides many others the $Sr_2RuO_4$ [4], Borides [5] and $NaCoO_2$:$H_2O$ [6] came through, the Fe based compounds [1,2] have taken the condensed matter physics community by surprise. Continuing the trend of observations of superconductivity in various layered structures, a week before successful synthesis of superconducting BiOCuS is reported by Ubaldini et al [7].

Interestingly the BiOCuS is not only iso-structural to by now famous REFeAsO, but also the ground state of these compounds. In fact the ground state stoichiometric REFeAsO is not yet available. It is believed that REFeAsO always have some oxygen deficiency and hence deviating possibly from its antiferromagnetic (AFM) ground state to the observed spin density wave (SDW). Host of this class of materials are recently reviewed by R. Pöttgen, D. Johrendt [8]. Although, the REFeAsO had became more popular due to its superconductivity and continued interest of condensed matter physicists, various other similar structure i.e., ZrCuSiAs-type compounds also received renewed attention [8]. One such example is the appearance of superconductivity in BiOCuS as reported by by Ubaldini etal [7]. This short cond-mat paper mentions the synthesis temperatures between 400 C and 800 C for 50 hours, and leaves a wide scope of speculation. More interestingly the Cu deficient $BiOCu_{1-x}S$ samples are reported superconducting with transition temperature as high as around 5.8K. If so, then undoubtedly story line is continuing from 1986 avalanche of Cu-O based cuprate superconductivity [3] to Fe based Fe-As [1] and Fe-Se [2] and now the Cu-S [7]. The striking similarity in all these compounds is that the superconductivity resides in their strongly correlated 3d metal based layer and other structural blocks provide the mobile carriers. Further their ground state is magnetic and insulating, though still there are some doubts about it, mainly because phase purity and stoichiometry of these materials are often in question. The Cu-O based high $T_c$ (HTSc) and Fe based Fe-As and Fe-Se superconductivity is well established, but the possible new entrant Cu-S is yet in question. So much so, that even after a week of the appearance of the first article [7], there is no confirmation of the $BiOCu_{1-x}S$ superconductivity yet. This is unlike HTSc and Fe based compounds, where it happened to be a day to day reporting of $T_c$ on popular con-mat site. Albeit a theoretical article on electronic band structure of BiOCuS as a parent phase for novel layered superconductors with possibility of superconductivity via appropriate doping has appeared [9]. In any case the prediction of superconductivity in CuS based layered compounds was predicted as early as in 2003 by Baskaran et al. [10]. If, the Cu-S based superconductivity proves right it will certainly be a feast to the theoreticians. The two decades old mystery of HTSc superconductivity may find its roots in iron-oxy arsenide/selenide [1,2] and oxy-copper sulfides [7].

To keep the ongoing interest alive, in the current article we report the phase formation temperature and the structural details of $BiOCu_{1-x}S$. Worth mentioning is the fact that the compound is already reported for its synthesis and some of physical properties in year 2008 [11]. Our interest here is to look for superconductivity via appropriate doping of carriers as done in case of Ubaldini et al. [7] and proposed theoretically in [9,10]. Although our preliminary studies did not approve the appearance of superconductivity in $BiOCu_{1-x}S$, but the current article is an effort to supplement the phase formation of BiOCuS [7,11] and possibility of superconductivity with appropriate doping of carriers. We also synthesized $Bi_{1-x}Pb_xOCuS$ (x = 0.02, 0.05) in single phase but not superconducting down to 4.2K.

Stoichiometric amounts of Bi, $Bi_2O_3$, Cu and S of better than 3N purity mixed in ratio of $1/3Bi_2O_3$ +0.34Bi +$Cu_{1-x}$ + S are pelletized and sealed in an evacuated quartz tube having vacuum of better than $10^{-3}$ Torr. The sealed quartz tubes containing various respective samples are heated at 500 $^0$C for 30 hours in a single step. Subsequently the samples are furnace cooled to room temperature. The heating/cooling rates are kept at $1^0$C/minute. Typically 1gram of raw pellet is sealed in 2.5 cm diameter and 10 cm length high quality quartz. The resultant compound taken out after breaking the quartz tubes are black in color and very brittle. In fact they are nearly powdered. The X-ray diffraction patterns of these compounds are taken on a Rigaku diffractometer. The Reitveld analysis is carried out on all the samples.

Fig.1 (a,b,c) depict the room temperature X-ray diifraction (XRD) patterns of $BiOCu_{1-x}S$. All the three studied samples are crystallized in tetragonal structure with P4/nmm space group. The The schematic crystal structure of BiOCuS is shown in Fig.2. The lattice parameters are a = 3.868(4) A and c = 8.557(4) A for x = 0.0 sample and the same is increased slightly for x = 0.10 and 0.15. The quality of fit parameters is mentioned on the XRD figures itself. Generally speaking the samples are phase pure with only minute amount of some unidentified species. Reitveld analysis is carried out for all the samples with in space group P4/nmm and Wyckoff positions as Bi(2c), O(2a), Cu(2b) and S(2c). The XRD Reitveld analysis fitted positions within mentioned scheme of the co-ordinates are Bi( ¼, ¼, 0.151), Cu(¾ , ¼, ½), S(¼ ¼, 0.648) and O(¾, ¼, 0) for BiOCuS. The electronic band structure calculations of BiOCuS resulted in Bi(z) as 0.1495 and S(z) as 0.664 [9]. The presently determined z coordinate positions for Bi and S from fitting of XRD patterns are in good agreement with theoretical values [9] and earlier reported experimental work [11]. As mentioned the BiOCuS structure and space group are same as that of REFeAsO. Comparing the two, it is seen that the Bi(z) is shifted to 0.15 from 0.13 for La in LaFeAsO [10], on the other hand the S(z) is nearly at same position as As(z) [12]. For Cu deficient samples i.e. with x = 0.10 (10%) and 0.15 (15%) the co-ordinates positions of Cu and O are fixed within the permitted space group and the z coordinates for Bi and S are respectively 0.151, 0.667 for the former and 0.151.and 0.657 for the later. The coordinate positions of Bi, O, Cu and S for all the BiOCu Samples are mentioned in Table 1. The lattice parameters of all the studied samples along with their unit cell volume are given in Table 2. The lattice parameters are close to that as reported in ref. 7.

As mentioned, the resultant compounds could not remain in pellet form and are powdered, hence the transport measurements including resistivity versus temperature could not be carried out on as synthesized samples. After successive two-step annealing (520 $^0$C for 12 hours) i.e. heating of re-pelletized powders, relatively dense and hard pellets are obtained which are good enough for the transport measurements. Against the prediction, none of the samples i.e. $BiOCu_{1-x}S$ with x = 0.0, 0.10 and 0.15 exhibited superconductivity transition down to 4.2 K. Albeit the room temperature conductivity improved by an order of magnitude for Cu deficient samples. The pristine sample i.e. BiOCuS is near insulating with high resistivity of around 100 Ohms-cm. We also synthesized single phase $Bi_{1-x}Pb_xOCuS$ (x = 0.02, 0.0.05), see Fig. 3 and found an improved conductivity of the samples by Bi site Pb doping but not superconducting, see Fig. 4. The order of the conductivity is same as reported in ref. 11. Samples with higher Pb content are yet to be realized in single phase. The physical properties of BiOCuS still need to be studied in detail, in particular the magnetic ground state (possibly AFM) of Cu and the effect of pressure on its transport properties. One will eventually be interested in BiOFeAs as well, i.e., if Bi site doping routes open up easily and the structure forms. None the less the hope of superconductivity in BiOCuS by Ubaldini et al [7] and nice band structure predictions [9] along with earlier theoretical inputs [10] are very welcome. *The doping routes via Bi-O, Tl-O and Hg-O layers to Cu-*$O_2$ *superconducting planes proved very fruitful in HTSc and could eventually be doing the same in case of FeAs and FeSe superconductors. (Bi/Tl/Hg)O-FeAs/FeSe may prove to be fruitful, provided the required ZrCuSiAs-type 1111 structure forms.* If any break through, our short cond-mat article could be cited. We are currently on task to look for these compounds.

Summarily we synthesized single phase BiOCuS samples with up to 15% Cu deficiency and also with Bi site up to 5% Pb doping via an easy solid state synthesis route. The Phase formation temperature is clearly identified. The structural details of the studied compound are provided as being determined by Reitveld analysis of the XRD patterns. As of now we have not been able to find superconductivity in $BiOCu_{1-x}S$ or $Bi_{1-x}Pb_xOCuS$ down to 4.2K, but efforts are underway to appropriately dope the Cu-S block and induce superconductivity in the same. Some suggestions to look for layered 1111 structure superconducting compounds are proposed.

**References**

1. Y. Kamihara, T. Watanabe, M. Hirano and H. Hosono, J. Am. Chem. Soc. 130, 3296 (2008).
2. F. C. Hsu, L. Y. Luo, K. W. Yeh, T. K. Chen, T. W. Huang and P. M. Wu, Proc. Natl. Acad. Sci. USA **105**, 14262 (2008).
3. J.G. Bednorz, K.A. Muller, Z. Phys. B. **64**, 189 (1986)
4. Y. Maeno, H. Hashimoto, K. Yoshida, S. Nishizaki, T. Fujita, J. G. Bednorz, L. Lichtenberg, Nature **372**, 532 (1994).
5. J. Nagamatsu, N. Nakagawa, T. Muranaka, Y. Zentani, and J. Akimitsu, Nature **410**, 63 (2001).
6. K. Takada, H. Sakurai, E. Takayama-Muromachi, F. Izumi, R.A. Dilanian, and T. Sasaki Nature **422**, 53 (2003).
7. A. Ubaldini, E. Giannini, C. Senatore and D. van der Marel, arXiv:0911.5305V2 [cond-mat.super-con] (2009)
8. R. Pöttgen, D. Johrendt, Z. Naturforsch. **63b**, 1135 (2008)
9. I.R. Shen and A.I. Ivanovskii, arXiv:0912.0416V2 [cond-mat.super-con] (2009).
10. G. Baskaran, Phys. Rev. Lett. **90**, 197007 (2003).
11. H. Hiramatsu, H. Yanagi, T. Kamiya, K. Ueda, M. Hirano and H. Hosono Chem., Mater., **20**, 326 (2008).
12. V.P.S. Awana, Arpita Vajpayee, Monika Mudgel, Anuj Kumar, R.S. Meena, Rahul Tripathi, Shiva Kumar, R.K. Kotnala, and H. Kishan. J. Superconductivity and Novel Magnetism, **21**, 167 (2008).

**Table 1.** Reitveld refine parameters for $BiOCu_{1-x}S$ Space group P4/nmm

| **Atom** | **Site** | **x** | **y** | **z** |
|---|---|---|---|---|
| Bi | 2c | 0.25 | 0.25 | 0.151(4) |
| Cu | 2b | 0.75 | 0.25 | 0.50 |
| S | 2c | 0.25 | 0.25 | 0.648(4) |
| O | 2a | 0.75 | 0.25 | 0.00 |

**Table 2.** Lattice parameters and cell volume of $BiOCu_{1-x}S$ (x = 0.0 - 0.15)

| **Sample** | **a (A )** | **c (A)** | **V ($A^3$)** | **Rp%** | **Rwp%** | **$chi^2$** |
|---|---|---|---|---|---|---|
| BiOCuS | 3.868(4) | 8.557(4) | 128.06 | 5.36 | 6.94 | 4.30 |
| $BiOCu_{0.9}S$ | 3.869(4) | 8.560(4) | 128.15 | 6.05 | 7.96 | 5.42 |
| $BiOCu_{0.85}S$ | 3.869(4) | 8.559(4) | 128.19 | 6.85 | 9.66 | 8.24 |

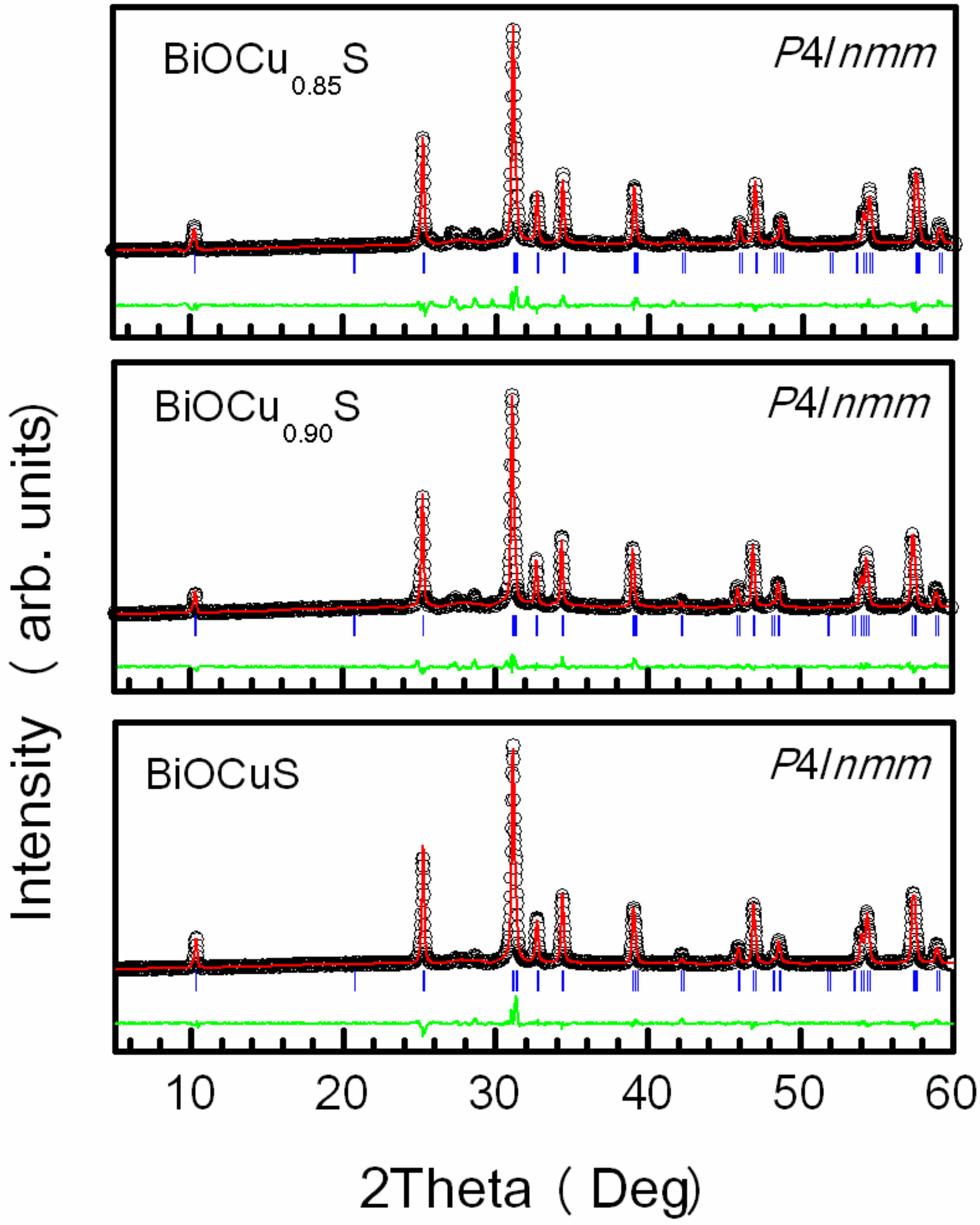

Figure 1. X-ray diffraction (XRD) patterns of $BiOCu_{1-x}S$( x = 0.0, 0.10, and 0.15).

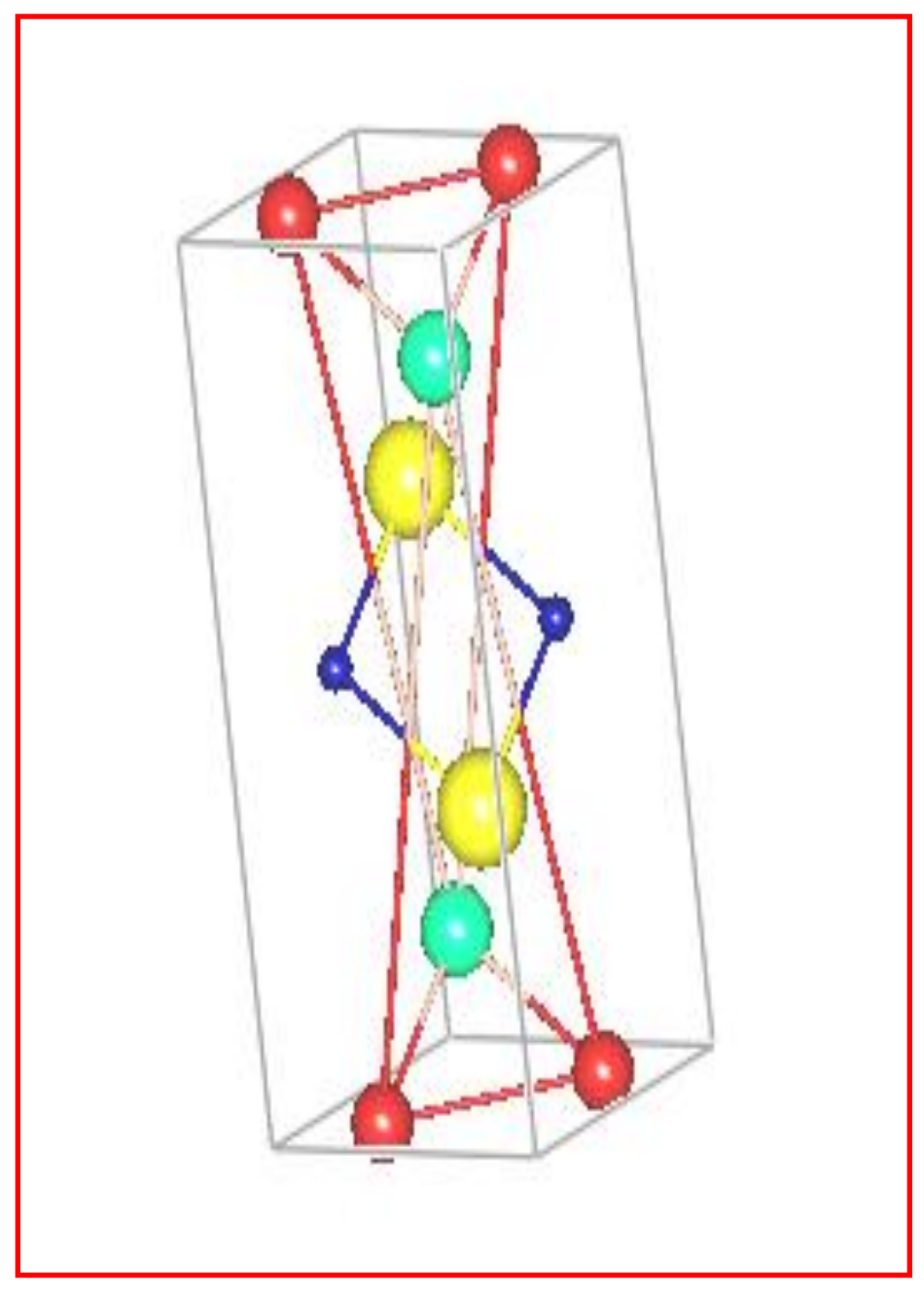

Figure 2. The schematic crystal structure of BiOCuS; the atoms are identified as Bi-green, O-red, Cu-blue and S-yellow.

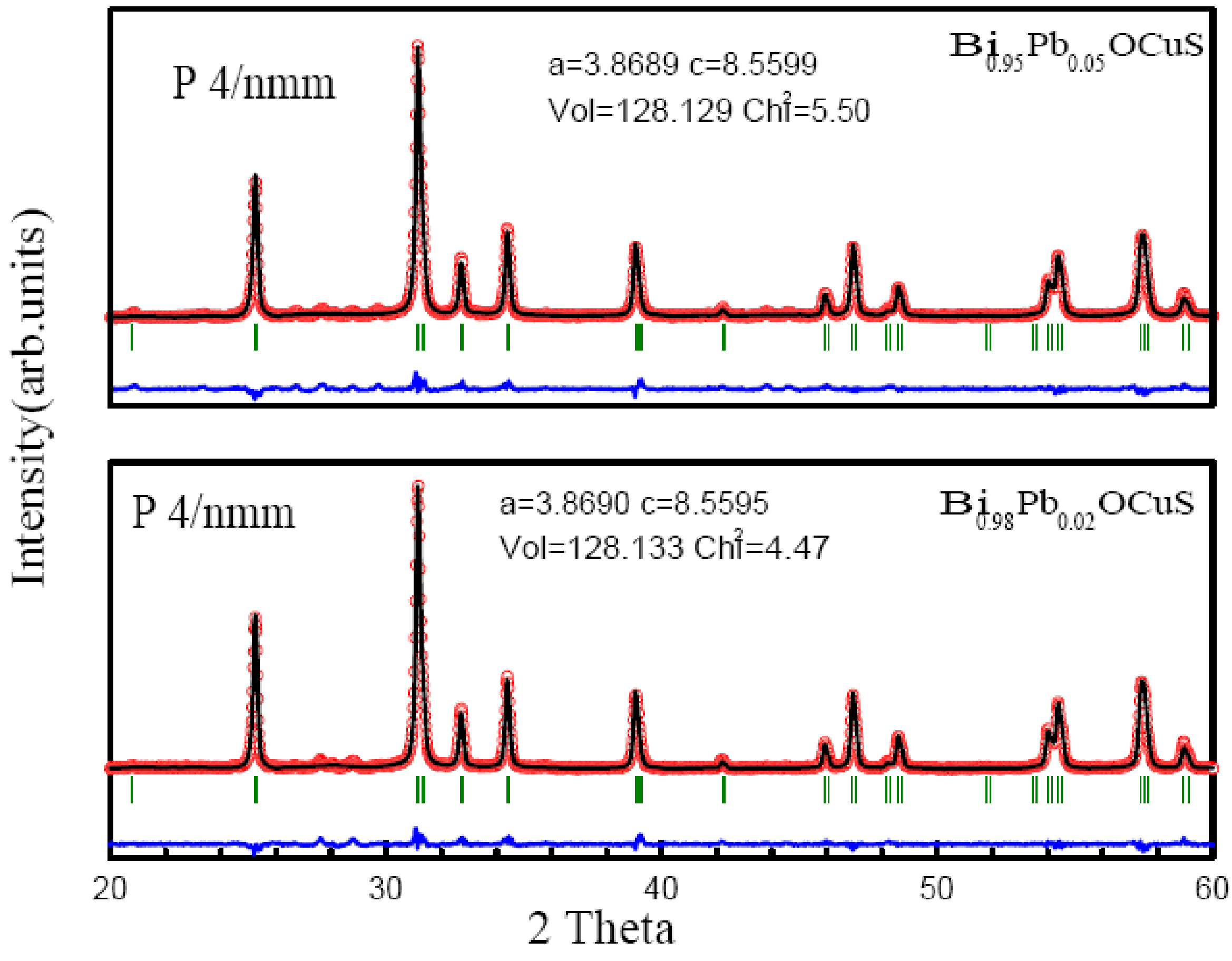

Figure 3. X-ray diffraction (XRD) patterns of $Bi_{1-x}Pb_xOCuS$ (x = 0.02, and 0.05).

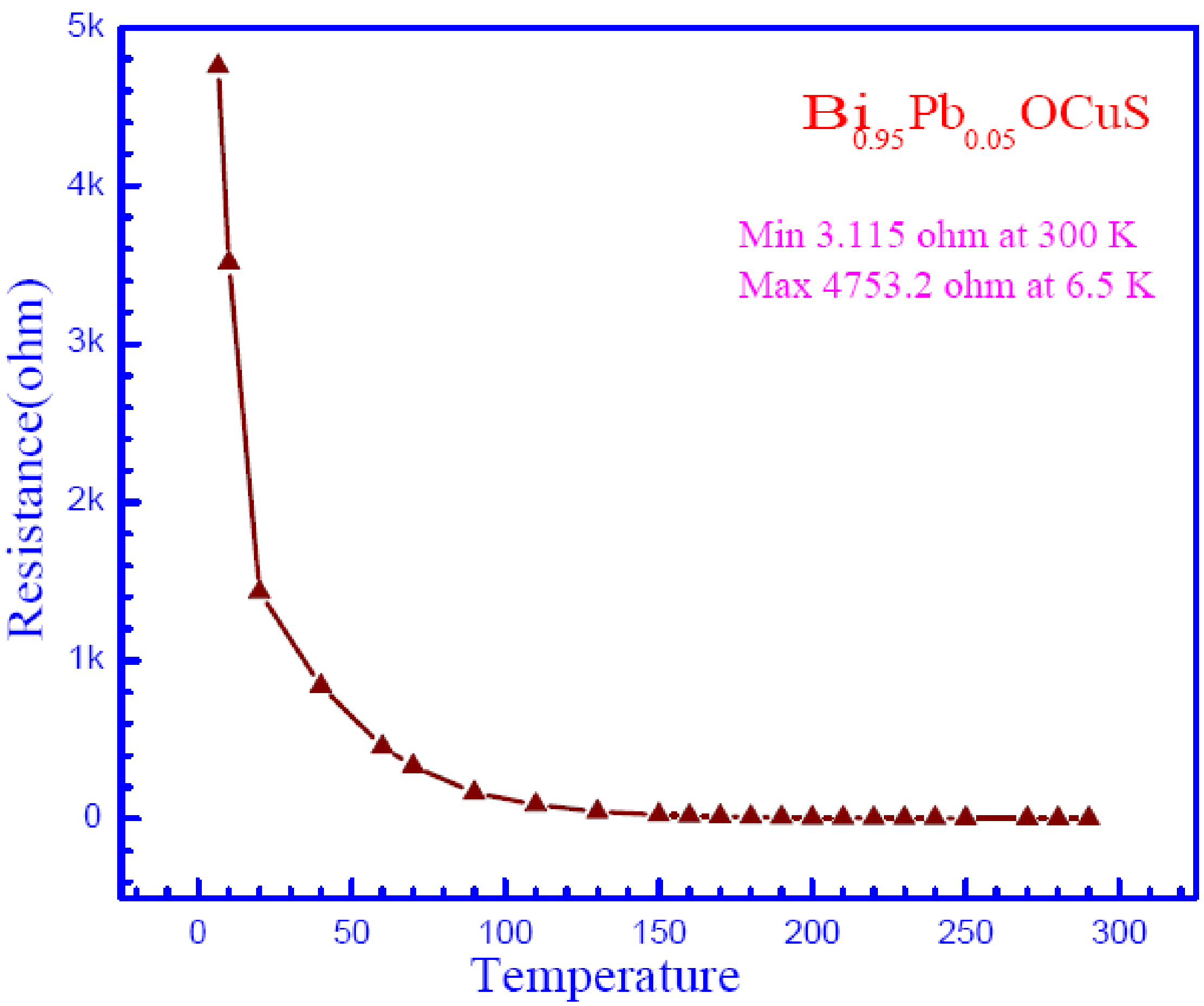

Figure 4. R(T) behavior of $Bi_{0.95}Pb_{0.05}CuS$